\documentclass[a4paper,11pt]{article}

\usepackage{lineno}
\usepackage{jheppub}
\usepackage{multirow}
\usepackage{array,hhline}
\usepackage{makecell}
\usepackage{hyperref}
\usepackage{booktabs}
\usepackage{floatrow}
\usepackage{autobreak}
\usepackage{longtable}
\usepackage{caption}
\usepackage{bbding}
\usepackage{bbm}
\usepackage{dsfont}
\usepackage{subfig}
\usepackage{pgffor}
\usepackage{forest}
\usetikzlibrary{shadows,arrows.meta}
\tikzset{
  parent/.style={align=center,text width=4cm,fill=gray!50,rounded corners=2pt},
  child/.style={align=center,text width=2.5cm,fill=gray!20,rounded corners=6pt},
  grandchild/.style={fill=white,text width=2.3cm}
}
\usepackage{listings}

\allowdisplaybreaks[4]

\newcommand{\ep}{\epsilon}
\newcommand\numberthis{\addtocounter{equation}{1}\tag{\theequation}}

\newcommand{\nn}{\nonumber \\}
\newcommand{\bff}{{\bf f}}
\newcommand{\bF}{{\bf F}}

\newcommand{\TTP}{Institute for Theoretical Particle Physics, KIT, 76128 Karlsruhe, Germany}

\title{\boldmath Three-loop ladder diagrams with two off-shell legs}

\author[]{Ming-Ming Long}

\affiliation[]{\TTP}
\preprint{TTP24-042, P3H-24-075}
\arxivnumber{2410.15431}

\emailAdd{ming-ming.long@kit.edu}

\abstract{
We present an analytic calculation of three-loop four-point Feynman integrals with two off-shell legs of equal mass.
We provide solutions to the canonical differential equations of two integral families in both Euclidean and physical regions. They are validated numerically against independent computations.
A total of 170 master integrals are expressed in terms of multiple polylogarithms up to weight six. 
Most of them are computed for the first time. Our results are essential ingredients of the scattering amplitudes for equal-mass diboson production at next-to-next-to-next-to-leading-order QCD at the LHC.
}

\begin{document}
\maketitle
\flushbottom

\section{Introduction}
\label{sec: intro}

Perturbative Quantum Chromodynamics (QCD) provides a powerful framework for understanding the physics of strong interactions at high energies. By systematically expanding in the strong coupling constant, predictions for observables such as cross sections and distributions in many scattering processes become increasingly accurate, largely due to significant advancements in the calculation of high-order QCD radiative corrections. 
Precise QCD calculations are critical for reducing theoretical uncertainties, thereby allowing for more stringent comparisons with experimental results. 
Achieving higher-order precision is essential not only for improving the accuracy of Standard Model (SM) predictions but also for probing potential new physics beyond the SM. 

At present, next-to-leading order (NLO) calculations have become the standard for scattering process predictions. However, moving to the next order, i.e., next-to-next-to-leading order (NNLO), introduces substantial complexities. One major challenge lies in the incomplete understanding of the function spaces that accommodate two-loop Feynman integrals (FIs)~\cite{Bourjaily:2022bwx}.  Nevertheless, in many cases, multi-loop FIs can be expressed in terms of multiple polylogarithms (MPLs)~\cite{Goncharov:1998kja, Goncharov:2001iea}. Specifically, in the context of dimensional regularization, the coefficients of the Laurent series expansion in the dimensional regulator $\ep$, introduced by the space-time dimensions $d = d_0 - 2\,\ep$~\footnote{Here, $d_0$ is an integer, typically 4 by default.}, can be expressed in terms of MPLs. Thanks to these insights and the development of numerous automatic tools, a wide range of important processes at the LHC are now understood at the NNLO level~\cite{Huss:2022ful}.

On one hand, pushing NNLO calculations to become the new standard remains an ambitious goal pursued by the community, although much progress is still required. On the other hand, there is a growing interest in exploring the feasibility of advancing beyond NNLO. It has been estimated that the upcoming High Luminosity upgrade of the LHC will necessitate calculations of many scattering processes at next-to-next-to-next-to-leading order (N3LO) to achieve precision at the percent level~\cite{Amoroso:2020lgh, Huss:2022ful, Caola:2022ayt}. In some specific cases, such as vector boson or Higgs boson production, N3LO predictions are already available~\cite{Heinrich:2020ybq}, but much of the N3LO landscape remains uncharted.

N3LO calculations face numerous challenges~\cite{Caola:2022ayt}, particularly in handling infrared and ultraviolet divergences that grow increasingly complex at higher orders. Techniques such as subtraction schemes, which are well-developed at NLO and NNLO, need further development for N3LO applications. 
Additionally, the computational demands of multi-loop FIs call for advanced algorithms and substantial resources~\cite{FebresCordero:2022psq}. Developing efficient techniques for tackling these challenges is critical to advancing QCD precision.

In the fully massless case, analytic results for four-point three-loop FIs have been obtained in Refs.~\cite{Henn:2013tua, Henn:2020lye}. These integrals have since been used to construct three-loop massless QCD scattering amplitudes, which are key for N3LO corrections to processes like diphoton, dijet, or photon-plus-jet production~\cite{Caola:2020dfu, Caola:2021rqz, Bargiela:2021wuy, Caola:2021izf, Caola:2022dfa, Bargiela:2022lxz}. The current state-of-the-art includes the calculation of four-point three-loop FIs with one off-shell external leg. While all the planar integrals can be expressed in terms of MPLs, the more intricate non-planar cases are still under active investigation~\cite{DiVita:2014pza, Canko:2021xmn, Canko:2023yoe, Henn:2023vbd, Gehrmann:2024tds}. 
These integrals are critical for computing scattering amplitudes relevant to important processes like Higgs-plus-jet~\cite{Gehrmann:2023etk} and $V+\text{jet}, (V=W, Z, \gamma^*)$ production~\cite{Gehrmann:2023zpz, Gehrmann:2023jyv} at the LHC.

In this paper, we take a further step by computing the four-point three-loop FIs with two off-shell legs of equal mass. 
At two loops, both planar and non-planar FIs have been available for several years~\cite{Gehrmann:2013cxs, Gehrmann:2014bfa}.
Very recently, the planar FIs with massive internal lines are computed in Ref.~\cite{He:2024iqg}.
These three-loop integrals contribute to the N3LO corrections to diboson production. 
Diboson production is a key process at the LHC, serving both as an important signal for precision tests of the electroweak sector and as a background for various new physics searches. Precise predictions for diboson production are crucial for ongoing Higgs boson studies, measurements of gauge boson couplings, and searches for anomalous interactions that could hint at physics beyond the SM. 

As a first step in this direction, we begin by considering the so-called three-loop ladder diagrams, as shown in Figure \ref{fig: diagrams}. The relevant kinematics and the definition of the two topologies are provided in the next section. In Section \ref{sec: deq}, we detail the construction of a canonical basis for the two integral families. By applying suitable boundary conditions, we derive solutions in terms of MPLs up to weight six in Section \ref{sec: soulution}. The conclusion and outlook are presented in Section \ref{sec: conclusion}. Additionally, an implementation of all the solutions is provided, with its usage explained in the appendix.
\section{Three-loop ladder diagrams with two off-shell legs}
\label{sec: family}

\begin{figure}[htbp]
  \centering
  \captionsetup[subfigure]{labelformat=empty}
  \subfloat[Ladder A, $\mathrm{LA}$]{%
    \includegraphics[width=0.45\textwidth]{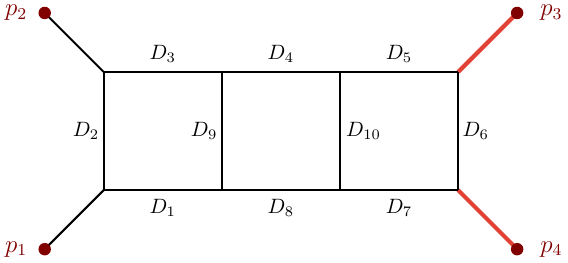}
  }
  \subfloat[Ladder B, $\mathrm{LB}$]{%
    \includegraphics[width=0.45\textwidth]{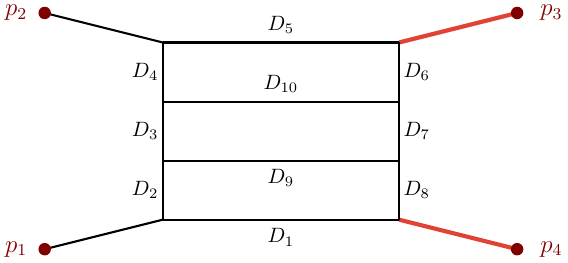}
  }
\caption{Two three-loop ladder topologies $\mathrm{LA}$ and $\mathrm{LB}$. The momenta $p_i$ are all incoming and the thick red lines represent the massive off-shell legs. The labels $D_i$ floating around the internal lines represent inverse propagators. See the text for their explicit definition.}
\label{fig: diagrams}
\end{figure}

We start in this section by specifying the definition of the topologies, or integral families, stemming from two three-loop ladder diagrams 
shown in Figure \ref{fig: diagrams}. 
The two families, dubbed Ladder A and B, for simplicity, will be called LA and LB in our later calculation. 
These two families are defined by the following two sets of inverse propagators~\footnote{Actually, the propagators of family $\mathrm{LB}$ are a 
re-ordering of the ones in $\mathrm{LA}$. However, we will treat them independently.},
\begin{itemize}
    \item \textbf{Family Ladder A} \\
    \begin{align}
	D_{1} &= l_1^2\,, & 
	D_{2} &= (l_1+p_1)^2\,, & 
	D_{3} &= (l_1+p_{12})^2\,, & 
	\nn
	D_{4} &= (l_2+p_{12})^2\,, & 
	D_{5} &= (l_3+p_{12})^2\,, & 
	D_{6} &= (l_3-p_4)^2\,, & 
	\nn
	D_{7} &= l_3^2\,, & 
	D_{8} &= l_2^2\,, & 
	D_{9} &= (l_1-l_2)^2\,, & 
	\numberthis  \label{eq:pros1} \\ 
	D_{10} &= (l_2-l_3)^2\,, & 
	D_{11} &= (l_1-p_4)^2\,, & 
	D_{12} &= (l_2-p_4)^2\,, & 
	\nn
	D_{13} &= (l_2+p_1)^2\,, & 
	D_{14} &= (l_3+p_1)^2\,, & 
	D_{15} &= (l_1-l_3)^2\,. & \nonumber
    \end{align}
    \item \textbf{Family Ladder B} \\
    \begin{align}
	D_1 &= l_1^2\,, & 
	D_2 &= (l_1+p_1)^2\,, & 
	D_{3} &= (l_2+p_1)^2\,, & 
	\nn
	D_{4} &= (l_3+p_1)^2\,, & 
	D_5 &= (l_3+p_{12})^2\,, & 
	D_6 &= (l_3-p_4)^2\,, & 
	\nn
	D_{7} &= (l_2-p_4)^2\,, & 
	D_{8} &= (l_1-p_4)^2\,, & 
	D_9 &= (l_1-l_2)^2\,, &
	\numberthis \label{eq:pros2} \\
	D_{10} &= (l_2-l_3)^2\,, & 
	D_{11} &= (l_1+p_{12})^2\,, & 
	D_{12} &= (l_2+p_{12})^2\,, & 
	\nn
	D_{13} &= l_3^2\,, & 
	D_{14} &= l_2^2\,, & 
	D_{15} &= (l_1-l_3)^2\,. & \nonumber
\end{align}
\end{itemize}
We used $p_{12}=p_1+p_2$ in above equations. The four external momenta satisfy the following kinematic relations,
\begin{equation}
    \sum p_i = 0, \quad p_{1,2}^2 = 0, \quad p_{3,4}^2 = m^2,
\end{equation}
where $m$ is the mass of two external off-shell legs. Three Mandenstam invariants $s,t,u$ are defined as
\begin{equation}
    s = (p_1 + p_2)^2, \quad 
    t = (p_1 + p_4)^2, \quad
    u = (p_2 + p_4)^2,
\end{equation}
and they are not dependent because
\begin{equation}
    s + t + u = 2\, m^2.
\end{equation}
In the center-of-mass frame of $p_1$ and $p_2$, two off-shell particles fly back to back. The invariants $t$ and $u$ can be written as
\begin{equation}
    t, u = m^2 -\frac{s}{2} \pm \frac{s}{2}\sqrt{1-\frac{4m^2}{s}} \cos\theta,
\end{equation}
where $\theta$ is the angle between the trajectory of $\Vec{p}_4$ and the beam line. In the physical region, $s$ should be above the threshold 
of the production of two particles with the same masses, i.e., $s > 4\, m^2$. And $t, u$ fulfill the following inequality,
\begin{equation}
   m^2 -\frac{s}{2} - \frac{s}{2}\sqrt{1-\frac{4m^2}{s}} < t, u < m^2 -\frac{s}{2} + \frac{s}{2}\sqrt{1-\frac{4m^2}{s}} < 0.
   \label{eq: kinematic constriant}
\end{equation}
We choose $s$ and $t$ as independent invariants. In the Euclidean region, we have
\begin{equation}
    s < 0, \quad t < 0, \quad m^2 <0.
\end{equation}
It is convenient to work with two dimensionless rations defined by
\begin{equation}
    v = \frac{s}{m^2}, \quad y = \frac{t}{m^2}.
\end{equation}
Multiplied by appropriate factors, the FIs can be normalized to be dimensionless and depend on $v, y$ only. We will first work in the non-physical region where
\begin{equation}
    s < 4m^2, \qquad m^2 < t < 0.
    \label{eq: nonphy region}
\end{equation}
It follows that
\begin{equation}
    v > 4, \qquad 0 < y <1.
\end{equation}
Then the results in the physical region could be obtained by analytic continuation according to Feynman's prescription which states that $s, t, m^2$ should
carry a positive vanishing imaginary part. 

We further introduce $x$ as follows,
\begin{equation}
    v = \frac{(1+x)^2}{x},
    \label{eq: Landau variable}
\end{equation}
which will be used to rationalize the square root in the canonical differential equation. As a result, in the
region defined by Eq. (\ref{eq: nonphy region}), we have
\begin{equation}
    0 < x < 1, \qquad 0< y < 1.
    \label{eq: nonphy region 2}
\end{equation}
While in the physical region, since $s$ and $m^2$ both change sign, $v$ is still positive and, therefore,
$x$ remains to stay between zero and unit. However, $y$ will become negative because only $m^2$ changes
its sign. It is important to be in the right branch when evaluating the solutions in the physical region. To do it, $y$ is understood to carry a positive infinitesimal imaginary part,
\begin{equation}
    y + i 0^+.
\end{equation}
Moreover, according to Eq. (\ref{eq: kinematic constriant}), $y$ should fulfill the following constraint,
\begin{equation}
    -\frac{1}{x} < y < -x.
    \label{eq: y constriant}
\end{equation}

\section{System of canonical differential equations}
\label{sec: deq}

The FIs we aim to compute have the following general form,
\begin{equation}
    F_{\Vec{a}} = \int \mathcal{D}^d l_1 \mathcal{D}^d l_2 \mathcal{D}^d l_3
    \frac{\prod_{j=11}^{15} D_j^{-a_j}}{\prod_{j=1}^{10} D_j^{a_j}}, \quad
    \begin{cases}
        a_j \in \mathbb{Z} & j \leq 10 \\
        a_j \in \mathbb{Z}^{\leq 0} & j > 10
    \end{cases},
    \label{eq:fi_definition}
\end{equation}
where $l_i$ are the loop momenta and the integration measure is defined as
\begin{equation}
    \mathcal{D}^d l_i = C_{\epsilon} \frac{(-m^2)^{\epsilon}}{i\pi^{d/2}} d^d l_i, \quad C_{\epsilon} = \frac{\Gamma (1-2\epsilon)}{\Gamma (1-\epsilon)^2 \Gamma (1+\epsilon)}.
\end{equation}
For each integral family, $\mathrm{LA}$ and $\mathrm{LB}$, the propagators in Eq. (\ref{eq:fi_definition}) take different forms, as shown in Eqs. (\ref{eq:pros1}) and (\ref{eq:pros2}). 
The FIs within each family are related through integration-by-parts (IBP) identities~\cite{Chetyrkin:1981qh, Laporta:2001dd}. Using \textsc{Kira}~\cite{Maierhoefer:2017hyi, Klappert:2020nbg} in combination with \textsc{FireFly}~\cite{Klappert:2019emp, Klappert:2020aqs} to perform IBP reduction, we identified 94 master integrals (MIs) for the $\mathrm{LA}$ family and 84 for the $\mathrm{LB}$ family. These MIs generally depend on the variables $s$, $t$, and $m^2$. A set of differential equations with respect to these three variables can be constructed in an automated manner~\cite{Lee:2012cn, Lee:2013mka}.
We treat the $\mathrm{LA}$ and $\mathrm{LB}$ families separately, although they share a few MIs, as will be demonstrated later.

Typically, the differential equations for a Laporta basis, generated by an IBP reducer, are highly complex. Transforming these equations into the canonical form~\cite{Henn:2013pwa} is often impractical using the available tools. A more feasible approach is to start with a better-chosen basis. Several strategies for this exist in the literature, and a recent summary of key principles applicable to many multi-loop FI calculations can be found in Ref.~\cite{Gorges:2023zgv}.

For integrals that can be factorized into products of lower-loop integrals, previous two-loop calculations provide useful guidance~\cite{Gehrmann:2013cxs, Gehrmann:2014bfa}. For genuinely three-loop integrals, some—typically from the lower sectors—have already been computed in the literature. However, the most challenging cases are the more complicated four-point integrals. To tackle this, we adopt a hybrid approach.

One method involves constructing the integrand in a special form, known as the $d$-log integrand~\cite{Henn:2020lye}, whose integration has a trivial leading singularity~\cite{Arkani-Hamed:2010pyv}. This makes it a promising candidate for forming a canonical basis, and such integrals are often referred to as "pure" in the literature. The \textsc{DlogBasis} package~\cite{Henn:2020lye} automates this process. However, in our case, this tool proved insufficient, as it only identified pure integrals in a few sectors~\footnote{\textsc{DlogBasis} was successful in computing the leading singularity for only a few sectors, and it encountered difficulties in sectors with more than seven propagators.}.

In the absence of a fully automated solution, we turn to experience-guided choices. This involves trying different MIs and analyzing the resulting differential equations, sector by sector. Starting with the simplest sector—those with the fewest propagators—we examine the block of differential equations corresponding to the integrals within that sector. This process is repeated iteratively, advancing to increasingly complex sectors until the top sector is reached.
At each iteration, we use a heuristic approach. For sectors with fewer than nine propagators and more than one MI, we favor MIs with double propagators over those with irreducible numerators. In sectors with many MIs, both types may appear. For the top sector and the next-to-top sector, however, irreducible numerators are preferred.
The double propagators and irreducible numerators are determined by trial and error.

Following the procedure outlined above, we obtain \textit{almost}-canonical differential equations that are linear in $\ep$. The non-canonical terms arise from dependencies of integrals in higher sectors on their respective sub-sectors. These can be systematically transformed into canonical form by applying the Magnus expansion~\cite{Argeri:2014qva}.
To present the canonical basis more clearly, we proceed in three steps for both integral families in the following subsections: \textit{i)} We start with an initial basis, a vector of integrals as defined in Eq. (\ref{eq:fi_definition}). \textit{ii)} A transformation that depends solely on $\ep$ is applied to this starting basis. \textit{iii)} Finally, the canonical basis is achieved through an additional transformation that depends only on the kinematics.

\subsection{Family $\mathrm{LA}$}
\begin{figure}[htbp]
\centering
\captionsetup[subfigure]{labelformat=empty}
\foreach \ii in {1,...,42}
{
\subfloat[$\mathcal{T}_{\ii}$]{%
    \includegraphics[width=0.15\textwidth]{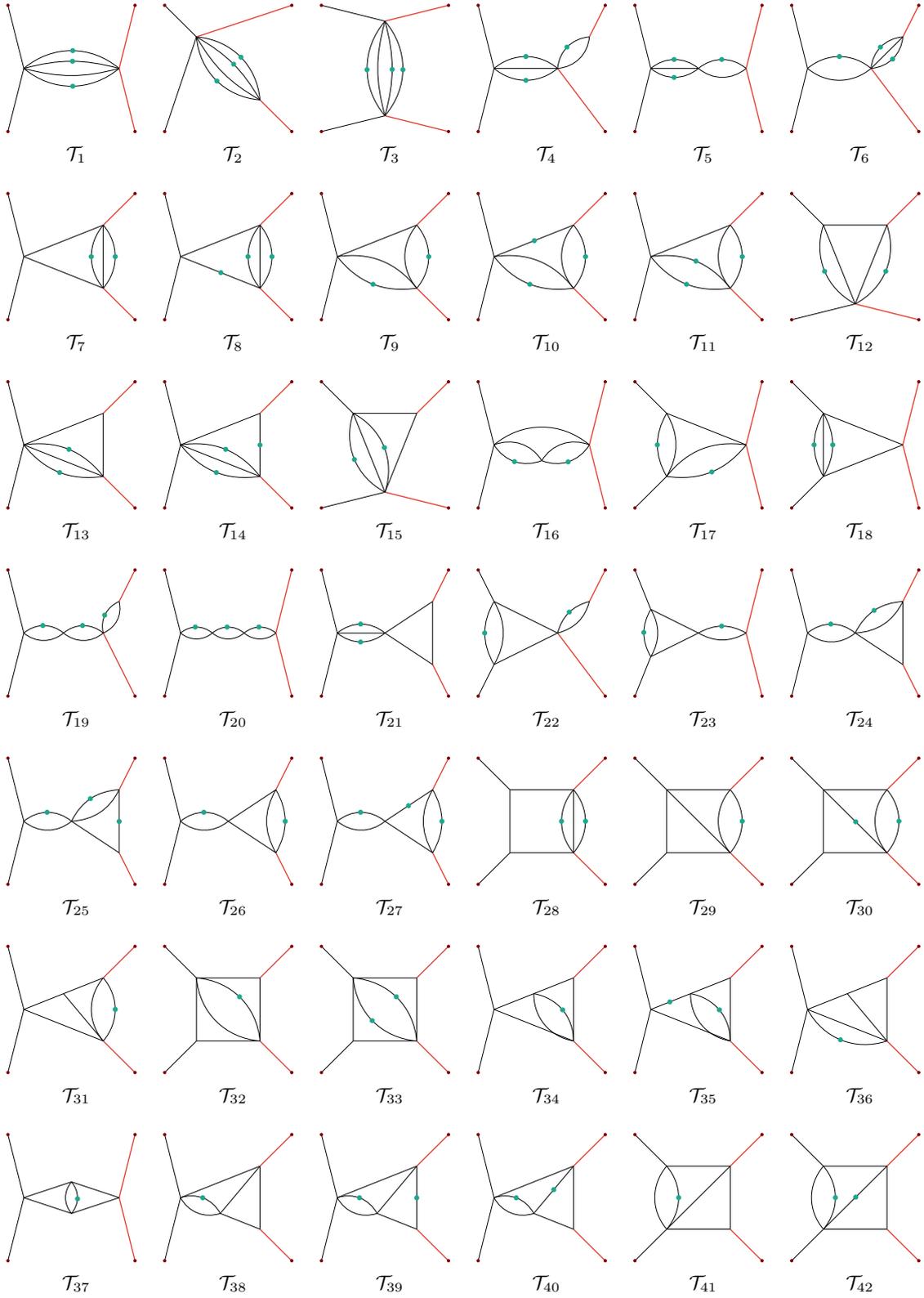}
  }
}
\caption{First part of MIs of the family $\mathrm{LA}$. Every green dot in one propagator means raising the corresponding power $a_j$ by one.}
\label{fig: F1misA}
\end{figure}

\begin{figure}[htbp]
\centering
\captionsetup[subfigure]{labelformat=empty}
\foreach \ii in {43,...,84}
{
\subfloat[$\mathcal{T}_{\ii}$]{%
    \includegraphics[width=0.15\textwidth]{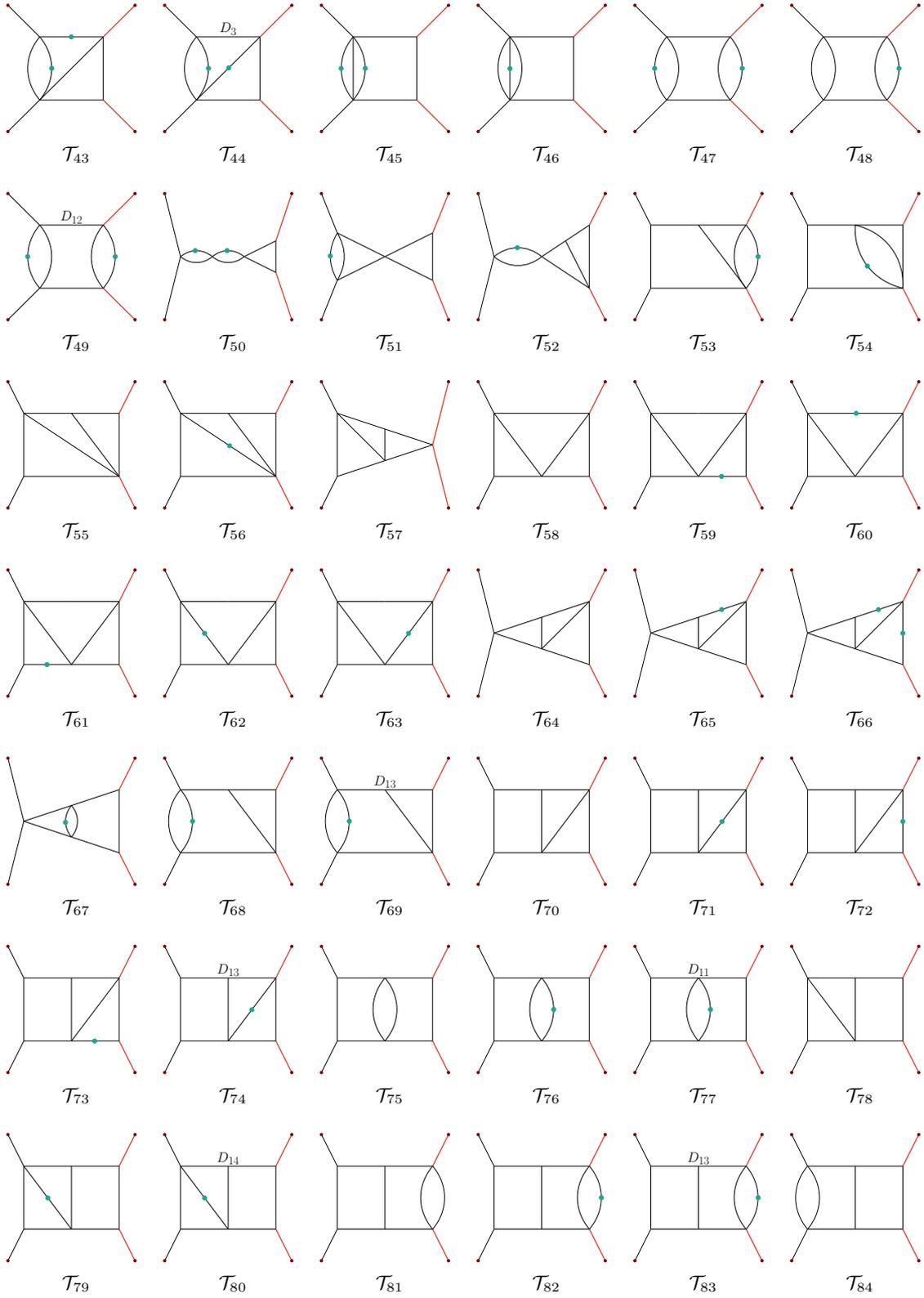}
  }
}
\caption{Second part of MIs of the family $\mathrm{LA}$. Every green dot in one propagator means raising the corresponding power $a_j$ by one.}
\label{fig: F1misB}
\end{figure}

\begin{figure}[htbp]
\centering
\captionsetup[subfigure]{labelformat=empty}
\foreach \ii in {85,...,94}
{
\subfloat[$\mathcal{T}_{\ii}$]{%
    \includegraphics[width=0.15\textwidth]{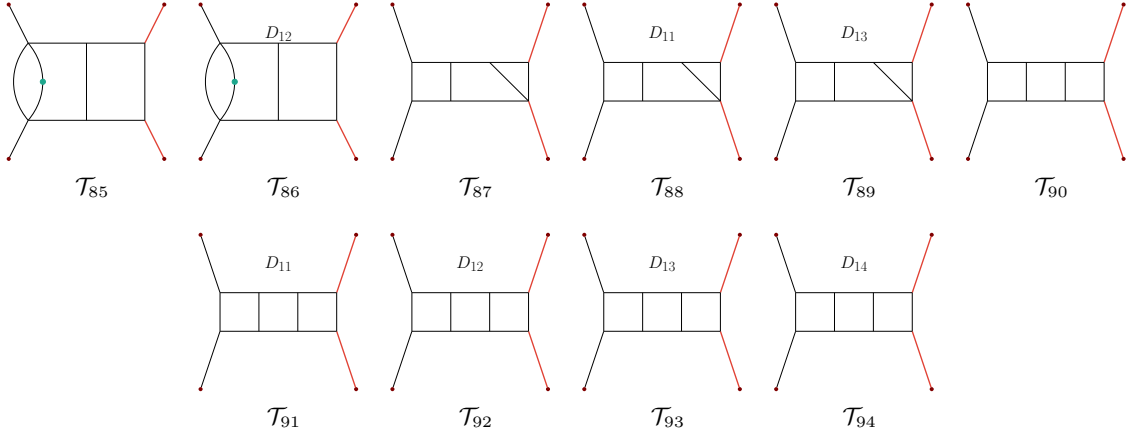}
  }
}
\caption{Last part of MIs of the family $\mathrm{LA}$. Every green dot in one propagator means raising the corresponding power $a_j$ by one.}
\label{fig: F1misC}
\end{figure}

We begin with the first integral family $\mathrm{LA}$. A starting basis for this family, $\mathcal{T}_{1,...,94}$, is given in Figures \ref{fig: F1misA}, \ref{fig: F1misB} and \ref{fig: F1misC}.
The differential equation system of this basis becomes much simpler than the Laporta one. The system is, however, neither linear nor canonical.
As we explained at the beginning of this section, we then apply a transformation depending on $\ep$ only as follows,
\begin{align}
    f_1 &= \mathcal{T}_1 \epsilon ^3 \, , &
f_2 &= \mathcal{T}_2 \epsilon ^3 \, , &
f_3 &= \mathcal{T}_3 \epsilon ^3 \, , &
f_4 &= \mathcal{T}_4 \epsilon ^3 \, , & \nn
f_5 &= \mathcal{T}_5 \epsilon ^3 \, , &
f_6 &= \mathcal{T}_6 \epsilon ^3 \, , &
f_7 &= \mathcal{T}_7 \epsilon ^4 \, , &
f_8 &= \mathcal{T}_8 \epsilon ^3 \, , & \nn
f_9 &= \mathcal{T}_9 \epsilon ^4 \, , &
f_{10} &= \mathcal{T}_{10} \epsilon ^3 \, , &
f_{11} &= \mathcal{T}_{11} \epsilon ^3 \, , &
f_{12} &= \mathcal{T}_{12} \epsilon ^4 \, , & \nn
f_{13} &= \mathcal{T}_{13} \epsilon ^4 \, , &
f_{14} &= \mathcal{T}_{14} \epsilon ^3 \, , &
f_{15} &= \mathcal{T}_{15} \epsilon ^4 \, , &
f_{16} &= \mathcal{T}_{16} \epsilon ^3 \beta_{+} \, , & \nn
f_{17} &= \mathcal{T}_{17} \epsilon ^4 \, , &
f_{18} &= \mathcal{T}_{18} \epsilon ^4 \, , & 
f_{19} &= \mathcal{T}_{19} \epsilon ^3 \, , &
f_{20} &= \mathcal{T}_{20} \epsilon ^3 \, , & \nn
f_{21} &= \mathcal{T}_{21} \epsilon ^4 \, , &
f_{22} &= \mathcal{T}_{22} \epsilon ^4 \, , &
f_{23} &= \mathcal{T}_{23} \epsilon ^4 \, , &
f_{24} &= \mathcal{T}_{24} \epsilon ^4 \, , & \nn
f_{25} &= \mathcal{T}_{25} \epsilon ^3 \, , &
f_{26} &= \mathcal{T}_{26} \epsilon ^4 \, , &
f_{27} &= \mathcal{T}_{27} \epsilon ^3 \, , &
f_{28} &= \mathcal{T}_{28} \epsilon ^4 \, , & \nn
f_{29} &= \mathcal{T}_{29} \epsilon ^5 \, , &
f_{30} &= \mathcal{T}_{30} \epsilon ^4 \, , & 
f_{31} &= \mathcal{T}_{31} \epsilon ^5 \, , &
f_{32} &= \mathcal{T}_{32} \epsilon ^5 \, , & \nn
f_{33} &= \mathcal{T}_{33} \epsilon ^4 \, , &
f_{34} &= \mathcal{T}_{34} \epsilon ^5 \, , &
f_{35} &= \mathcal{T}_{35} \epsilon ^3 \beta_{+} \, , &
f_{36} &= \mathcal{T}_{36} \epsilon ^5 \, , & \nn
f_{37} &= \mathcal{T}_{37} \beta_{-} \epsilon ^4 \, , &
f_{38} &= \mathcal{T}_{38} \epsilon ^5 \, , &
f_{39} &= \mathcal{T}_{39} \epsilon ^4 \, , &
f_{40} &= \mathcal{T}_{40} \epsilon ^3 \beta_{+} \, , & \nn
f_{41} &= \mathcal{T}_{41} \epsilon ^5 \, , &
f_{42} &= \mathcal{T}_{42} \epsilon ^4 \, , & 
f_{43} &= \mathcal{T}_{43} \epsilon ^3 \beta_{+} \, , &
f_{44} &= \frac{\mathcal{T}_{44} \beta_{-} \epsilon ^4}{\alpha_{-}} \, , & \nn
f_{45} &= \mathcal{T}_{45} \epsilon ^4 \, , &
f_{46} &= \mathcal{T}_{46} \gamma_{-} \epsilon ^4 \, , &
f_{47} &= \mathcal{T}_{47} \epsilon ^4 \, , &
f_{48} &= \mathcal{T}_{48} \beta_{-} \epsilon ^4 \, , & \nn
f_{49} &= \mathcal{T}_{49} \epsilon ^4 \, , &
f_{50} &= \mathcal{T}_{50} \epsilon ^4 \, , &
f_{51} &= \mathcal{T}_{51} \epsilon ^5 \, , &
f_{52} &= \mathcal{T}_{52} \epsilon ^5 \, , & \nn
f_{53} &= \mathcal{T}_{53} \epsilon ^5 \, , &
f_{54} &= \mathcal{T}_{54} \epsilon ^5 \, , &
f_{55} &= \mathcal{T}_{55} \epsilon ^6 \, , &
f_{56} &= \mathcal{T}_{56} \epsilon ^5 \, , & \nn
f_{57} &= \mathcal{T}_{57} \epsilon ^6 \, , &
f_{58} &= \mathcal{T}_{58} \epsilon ^6 \, , &
f_{59} &= \mathcal{T}_{59} \epsilon ^5 \, , &
f_{60} &= \mathcal{T}_{60} \epsilon ^5 \, , & \nn
f_{61} &= \mathcal{T}_{61} \epsilon ^4 \alpha_{+} \, , &
f_{62} &= \mathcal{T}_{62} \epsilon ^5 \, , &
f_{63} &= \mathcal{T}_{63} \epsilon ^5 \, , &
f_{64} &= \mathcal{T}_{64} \epsilon ^6 \, , & \nn
f_{65} &= \mathcal{T}_{65} \epsilon ^4 \alpha_{+} \, , &
f_{66} &= \mathcal{T}_{66} \epsilon ^3 \alpha_{+} \, , & 
f_{67} &= \mathcal{T}_{67} \epsilon ^5 \, , &
f_{68} &= \mathcal{T}_{68} \epsilon ^5 \, , & \nn
f_{69} &= \mathcal{T}_{69} \epsilon ^5 \, , &
f_{70} &= \mathcal{T}_{70} \epsilon ^6 \, , &
f_{71} &= \mathcal{T}_{71} \epsilon ^5 \, , &
f_{72} &= \mathcal{T}_{72} \epsilon ^5 \, , & \nn
f_{73} &= \mathcal{T}_{73} \epsilon ^5 \, , &
f_{74} &= \mathcal{T}_{74} \epsilon ^5 \, , &
f_{75} &= \mathcal{T}_{75} \beta_{-} \epsilon ^5 \, , &
f_{76} &= \mathcal{T}_{76} \epsilon ^5 \, , & \nn
f_{77} &= \mathcal{T}_{77} \epsilon ^5 \, , &
f_{78} &= \mathcal{T}_{78} \epsilon ^6 \, , &
f_{79} &= \mathcal{T}_{79} \epsilon ^5 \, , &
f_{80} &= \mathcal{T}_{80} \epsilon ^5 \, , & \nn
f_{81} &= \mathcal{T}_{81} \beta_{-} \epsilon ^5 \, , &
f_{82} &= \mathcal{T}_{82} \epsilon ^5 \, , &
f_{83} &= \mathcal{T}_{83} \epsilon ^5 \, , &
f_{84} &= \mathcal{T}_{84} \beta_{-} \epsilon ^5 \, , & \nn
f_{85} &= \mathcal{T}_{85} \epsilon ^5 \, , &
f_{86} &= \mathcal{T}_{86} \epsilon ^5 \, , &
f_{87} &= \mathcal{T}_{87} \epsilon ^6 \, , &
f_{88} &= \mathcal{T}_{88} \epsilon ^6 \, , & \nn
f_{89} &= \mathcal{T}_{89} \epsilon ^6 \, , &
f_{90} &= \mathcal{T}_{90} \epsilon ^6 \, , &
f_{91} &= \mathcal{T}_{91} \epsilon ^6 \, , &
f_{92} &= \mathcal{T}_{92} \epsilon ^6 \, , & \nn
f_{93} &= \mathcal{T}_{93} \epsilon ^6 \, , &
f_{94} &= \mathcal{T}_{94} \epsilon ^6 \, , &
\label{eq:linear basis 1}
\end{align}
where
\begin{equation}
    \alpha_{\pm} = 1 \pm \ep, \quad \beta_{\pm} = 1 \pm 2\ep, \quad \gamma_{\pm} = 1 \pm 3\ep.
\end{equation}
It is straightforward to verify that the new basis $\bff$, defined by Eq. (\ref{eq:linear basis 1}), satisfies a linear differential equation. After applying a transformation that depends solely on kinematic variables—determined either through a leading-singularity analysis or via a Magnus expansion~\footnote{We also found that the entire linear system can be transformed using only the Magnus expansion. This approach is similarly applicable to the integral family LB.}—we are able to obtain a canonical basis~\footnote{Recently, it was reported in Ref.~\cite{Jiang:2024eaj} that a canonical basis was obtained by a different approach, though no explicit solution was provided.} $\bF$ given by
\begin{align}
    F_1 &= f_1 s \, , &
F_2 &= f_2 m^2 \, , &
F_3 &= f_3 t \, , &  \nn
F_4 &= f_4 m^2 s \, , &
F_5 &= f_5 s^2 \, , &
F_6 &= f_6 m^2 s \, , &  \nn
F_7 &= f_7 \lambda \, , &
F_8 &= f_8 m^2 s-2 f_7 s \, , &
F_9 &= f_9 \lambda \, , & \nn
F_{10} &= f_{10} m^2 s-2 f_9 s \, , & 
F_{11} &= f_{11} m^2 s-4 f_9 s \, , &
F_{12} &= f_{12} \left(m^2-t\right) \, , & \nn
F_{13} &= f_{13} \lambda \, , &
F_{14} &= f_{14} m^4+2f_{13} \left(s-2 m^2\right) \, , &
F_{15} &= f_{15} \left(m^2-t\right) \, , & \nn
F_{16} &= f_{16} s \, , &
F_{17} &= f_{17} s \, , &
F_{18} &= f_{18} s \, , & \nn
F_{19} &= f_{19} m^2 s^2 \, , &
F_{20} &= f_{20} s^3 \, , &
F_{21} &= f_{21} \lambda  s \, , & \nn
F_{22} &= f_{22} m^2 s \, , &
F_{23} &= f_{23} s^2 \, , &
F_{24} &= f_{24} \lambda  s \, , & \nn
F_{25} &= 
\rlap{$\displaystyle
f_{25} m^4 s+\frac{3}{2} f_{24} s \left(s-2 m^2\right) \, ,
F_{26} = f_{26} \lambda  s \, ,
F_{27} = f_{27} m^2 s^2-\frac{3 f_{26} s^2}{2} \, ,
$}
 & \nn
F_{28} &= f_{28} s t \, , &
F_{29} &= f_{29} \left(m^2-s-t\right) \, , &
F_{30} &= f_{30} s t \, , & \nn
F_{31} &= f_{31} \lambda \, , &
F_{32} &= f_{32} \left(m^2-s-t\right) \, , &
F_{33} &= f_{33} s t \, , & \nn
F_{34} &= f_{34} \lambda \, , & \nn
F_{35} &= 
\rlap{$\displaystyle
f_{35} m^2 s
-3f_{34} \left(m^2+\frac{3 s}{2}\right) 
+\frac{3}{4} f_{13} \left(s-2 m^2\right)
+\frac{1}{4} f_7 \left(2 m^2-s\right)
\, ,$} & \nn
F_{36} &= f_{36} \lambda \, , &
F_{37} &= f_{37} s \, , &
F_{38} &= f_{38} \lambda \, , & \nn
F_{39} &= f_{39} \lambda  m^2 \, , &
F_{40} &= f_{40} m^2 s+\frac{3}{2} f_{39} m^2 s-4 f_9 s \, , &
F_{41} &= f_{41} \left(m^2-s-t\right) \, , & \nn
F_{42} &= f_{42} s t \, , &
F_{43} &= f_{43} s \left(m^2-t\right)-12 f_{41} \left(m^2-t\right) \, , &
F_{44} &= f_{44} \lambda -f_{15} \lambda \, , & \nn
F_{45} &= f_{45} s t \, , &
F_{46} &= f_{46} \lambda \, , &
F_{47} &= f_{47} s t \, , & \nn
F_{48} &= f_{48} \lambda \, , &
F_{49} &= f_{49} s \, , &
F_{50} &= f_{50} \lambda  s^2 \, , & \nn
F_{51} &= f_{51} \lambda  s \, , &
F_{52} &= f_{52} \lambda  s \, , &
F_{53} &= f_{53} s \left(m^2-t\right) \, , & \nn
F_{54} &= f_{54} s \left(m^2-t\right) \, , &
F_{55} &= f_{55} \left(m^2-s-t\right) \, , &
F_{56} &= f_{56} s \left(m^2-t\right) \, , & \nn
F_{57} &= f_{57} s \, , &
F_{58} &= f_{58} \left(m^2-t\right) \, , &
F_{59} &= f_{59} m^2 s \, , & \nn
F_{60} &= f_{60} \lambda  s \, , &
F_{61} &= f_{61} s \left(m^2-t\right) \, , &
F_{62} &= f_{62} s \left(m^2-t\right) \, , & \nn
F_{63} &= f_{63} s t \, , &
F_{64} &= f_{64} \lambda \, , &
F_{65} &= f_{65} \lambda  s \, , & \nn
F_{66} &=  
\rlap{$\displaystyle
f_{66}m^4 s
-f_{65} s \left(m^2-2 s\right)
-2 f_{64} \left(m^2+4 s\right)
+3 f_{39} m^2 s
-2 f_{38} \left(m^2+4 s\right)
+2f_{34} \left(2 m^2- s\right)
$} & \nn
& \quad \rlap{$\displaystyle
+f_{31} \left(m^2-8 s\right)
+f_{24} s \left(s-2 m^2\right)
+\frac{3 f_{13} m^2}{2} 
+f_9 \left(m^2-4 s\right)
-f_7 \left(m^2+s\right)
\, ,
$} & \nn
F_{67} &= f_{67} \lambda  s \, , &
F_{68} &= f_{68} s \left(m^2-t\right) \, , &
F_{69} &= f_{69} \lambda \, , & \nn
F_{70} &= f_{70} s \left(s+t-m^2\right) \, , &
F_{71} &= f_{71} s^2 t \, , &
F_{72} &= f_{72} m^2 s \left(m^2-t\right) \, , & \nn
F_{73} &= f_{73} m^2 s^2 \, , &
F_{74} &= f_{74} \lambda  s \, , &
F_{75} &= f_{75} \lambda  s \, , & \nn
F_{76} &= f_{76} s^2 t \, , &
F_{77} &= f_{77} s^2+f_{45} m^2 s-\frac{1}{3} f_{33} m^2 s \, , &
F_{78} &= f_{78} s \left(s+t-2 m^2\right) \, , & \nn
F_{79} &= f_{79} s^2 t \, , &
F_{80} &= f_{80} \lambda  s \, , &
F_{81} &= f_{81} s^2 \, , & \nn
F_{82} &= f_{82} s^2 t \, , &
F_{83} &= f_{83} \lambda  s \, , &
F_{84} &= f_{84} \lambda  s \, , & \nn
F_{85} &= f_{85} s^2 t \, , &
F_{86} &= f_{86} s^2+f_{45} m^2 s-\frac{2}{3} f_{42} m^2 s \, , &
F_{87} &= f_{87} s^2 \left(m^2-t\right) \, , & \nn
F_{88} &= f_{88} s^2-f_{87} m^2 s^2 \, , &
F_{89} &= f_{89} \lambda  s \, , &
F_{90} &= f_{90} s^3 t \, , & \nn
F_{91} &= 
\rlap{$\displaystyle
f_{91} s^3 
+2 f_{85} m^2 s^2
-f_{79} m^2 s^2
\, , 
F_{92} = 
f_{92} s^3 
-f_{79} m^2 s^2
+2 f_{76} m^2 s^2
-f_{71} m^2 s^2
\, , 
$} & \nn
F_{93} &= f_{93} \lambda  s^2 \, , &
F_{94} &= f_{94} \lambda  s^2 \, , &
\label{eq: ut basis 1}
\end{align}
with $\lambda$ the square root
\begin{equation}
  \lambda = \sqrt{s(s-4m^2)}.
\end{equation}
As we said, the transformation in Eq. (\ref{eq: Landau variable}) can rationalize $\lambda$,
\begin{equation}
    \lambda^2 = (m^2)^2 \frac{(1-x^2)^2}{x^2}.
\end{equation}
And note that the canonical basis $\bF$ has been normalized to be dimensionless, depending on $x, y$ only. It satisfies two differential equations
of the form
\begin{equation}
    \partial_x \bF = \epsilon \mathrm{A}_x \bF, \quad \partial_y \bF = \epsilon \mathrm{A}_y \bF,
\end{equation}
which can be recast into the $d$log form
\begin{equation}
    {\rm d} \bF = \epsilon {\rm d}\mathbb{A} \bF.
    \label{eq: dlog}
\end{equation}
The matrix $\mathbb{A}$ are composed of 9 logarithms
\begin{equation}
    \mathbb{A} = \sum_{i=1}^{9} \mathbb{C}_i \log(\omega_i),
\end{equation}
where $\mathbb{C}_i$ are constant matrices and $\omega_i$ are called symbol letters (or simply letters) and they read
\begin{align}
\label{eq: letters1}
\omega _1 &= x\, ,&
\omega _4 &= 1+x\, ,&
\omega _7 &= 1+x y\, ,& \nn
\omega _2 &= y\, ,&
\omega _5 &= 1-y\, ,&
\omega _8 &= 1+ x y +x^2\, ,& \\
\omega _3 &= 1-x\, ,&
\omega _6 &= x+y\, ,&
\omega _9 &= 1+x+x y+x^2\, .& \nonumber
\end{align}
No new letters appear compared to the two-loop case~\cite{Gehrmann:2013cxs, Gehrmann:2014bfa}. 
It is clear that in the region defined by Eq. (\ref{eq: nonphy region}), every letter is positive. 
However, in the physical region, $\omega_{2, 6}$ are negative.

\subsection{Family $\mathrm{LB}$}

\begin{figure}[htbp]
\centering
\captionsetup[subfigure]{labelformat=empty}
\foreach \ii in {1,...,42}
{
\subfloat[$\mathcal{T}_{\ii}$]{%
    \includegraphics[width=0.15\textwidth]{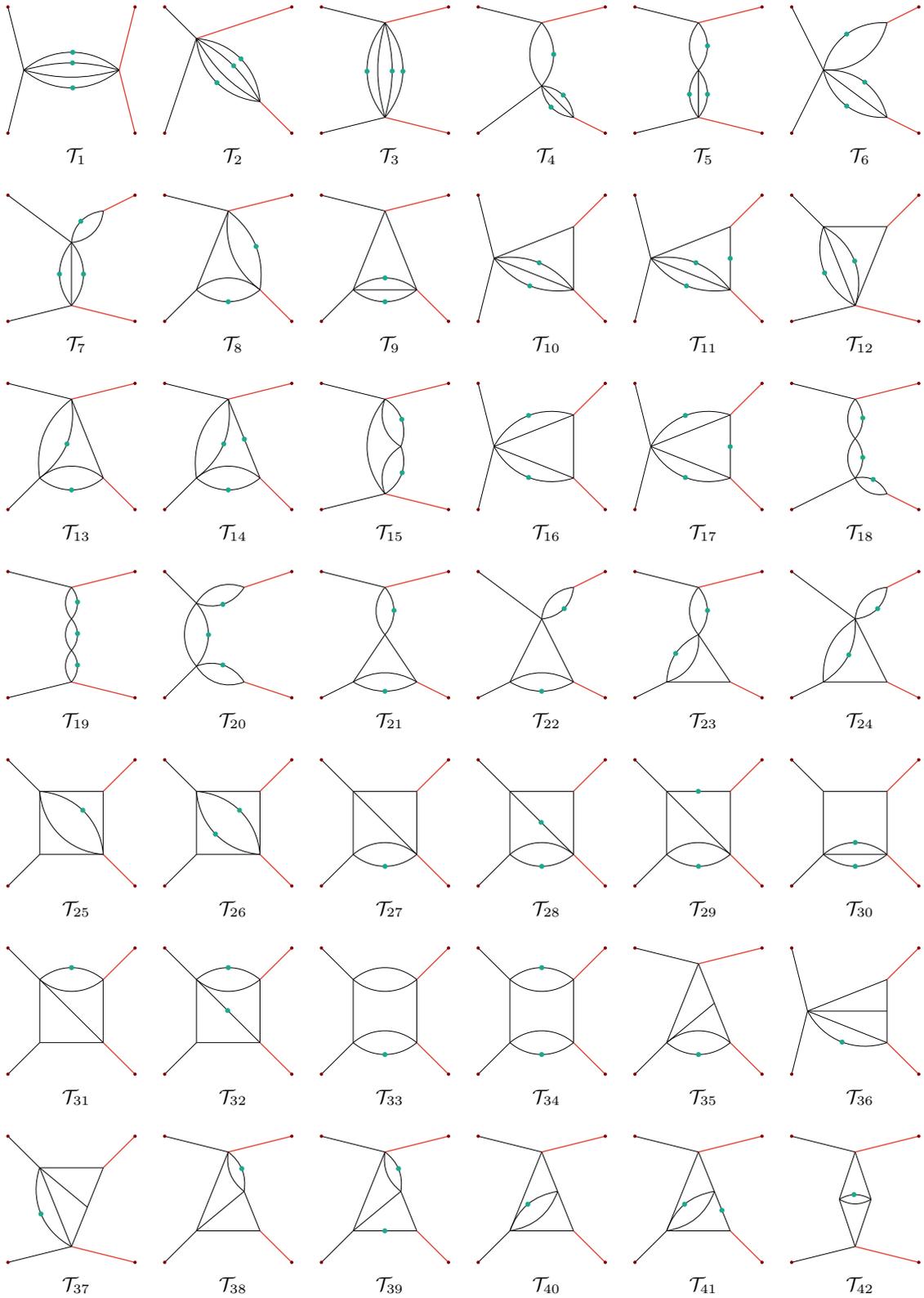}
  }
}
\caption{First part of MIs of the family $\mathrm{LB}$. Every green dot in one propagator means raising the corresponding power $a_j$ by one.}
\label{fig: F2misA}
\end{figure}

\begin{figure}[htbp]
\centering
\captionsetup[subfigure]{labelformat=empty}
\foreach \ii in {43,...,84}
{
\subfloat[$\mathcal{T}_{\ii}$]{%
    \includegraphics[width=0.15\textwidth]{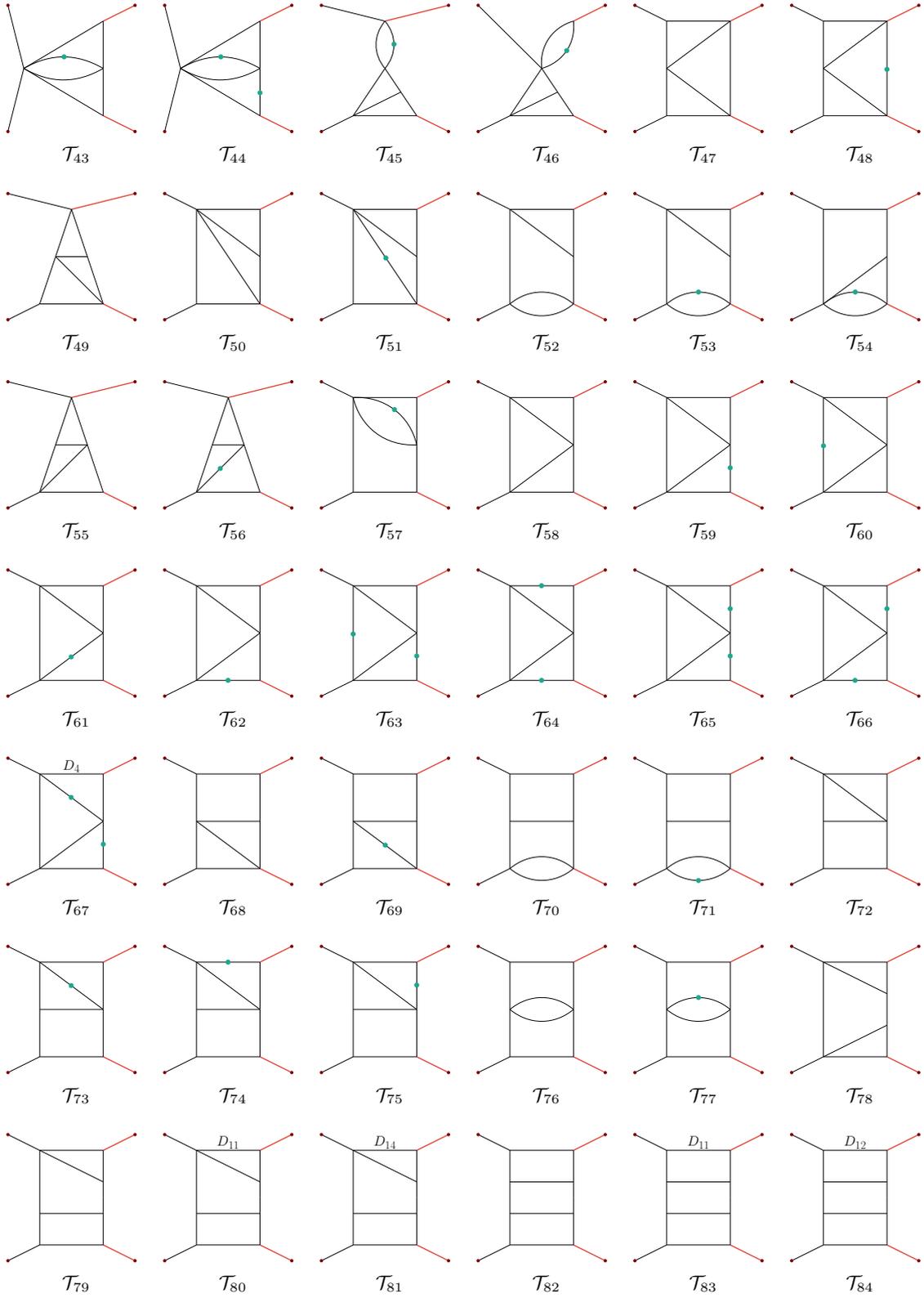}
  }
}
\caption{Second part of MIs of the family $\mathrm{LB}$. Every green dot in one propagator means raising the corresponding power $a_j$ by one.}
\label{fig: F2misB}
\end{figure}

We now turn to the second integral family, denoted as $\mathrm{LB}$. The procedure follows a similar approach to that of the previous subsection. First, the starting basis for family $\mathrm{LB}$ is presented in Figures \ref{fig: F2misA} and \ref{fig: F2misB}.
Notably, integrals $\mathcal{T}_{1,2,3,10,11,12,25,26}$ have already appeared in family $\mathrm{LA}$, corresponding to integrals $\mathcal{T}_{1,2,3,13,14,15,32,33}$ in Figure \ref{fig: F1misA}. Given that most of the MIs in family $\mathrm{LB}$ are new, we opted to compute them independently of the ones in the previous subsection.

The selection of the starting basis for family $\mathrm{LB}$ is more intricate than for family $\mathrm{LA}$. This added complexity arises from the fact that some sectors contain a significantly higher number of MIs. In family $\mathrm{LA}$, the maximum number of MIs in any sector is six, which includes the sector containing MI $\mathcal{T}_{58}$. However, in family $\mathrm{LB}$, this number increases to ten, notably in the sector containing MI $\mathcal{T}_{58}$, as shown in Figure \ref{fig: F2misB}.
Additionally, a clear symmetry can be observed in the diagram representation of $\mathcal{T}_{58}$ when flipped vertically. This symmetry restricts the possible choices of MIs, increasing the complexity of selecting candidate integrals~\footnote{Due to this symmetry, there are only four independent ways to introduce a double propagator, despite there being seven propagators in the denominator. For configurations involving two double propagators, the number of possibilities increases considerably.}.
After extensive testing of different combinations, we selected four candidate MIs with one double propagator and four with two double propagators. The final and most complex candidate MI contains two double propagators and an irreducible numerator.

Using the starting basis, we apply a transformation that depends only on $\epsilon$, as defined in the following equations,
\begin{align}
    f_1 &= \mathcal{T}_1 \epsilon ^3 \, , &
f_2 &= \mathcal{T}_2 \epsilon ^3 \, , &
f_3 &= \mathcal{T}_3 \epsilon ^3 \, , &
f_4 &= \mathcal{T}_4 \epsilon ^3 \, , & \nn
f_5 &= \mathcal{T}_5 \epsilon ^3 \, , &
f_6 &= \mathcal{T}_6 \epsilon ^3 \, , &
f_7 &= \mathcal{T}_7 \epsilon ^3 \, , &
f_8 &= \mathcal{T}_8 \epsilon ^4 \, , & \nn
f_9 &= \mathcal{T}_9 \epsilon ^4 \, , &
f_{10} &= \mathcal{T}_{10} \epsilon ^4 \, , &
f_{11} &= \mathcal{T}_{11} \epsilon ^3 \, , &
f_{12} &= \mathcal{T}_{12} \epsilon ^4 \, , & \nn
f_{13} &= \mathcal{T}_{13} \epsilon ^4 \, , &
f_{14} &= \mathcal{T}_{14} \epsilon ^3 \, , &
f_{15} &= \mathcal{T}_{15} \epsilon ^3 \beta_{+} \, , &
f_{16} &= \mathcal{T}_{16} \epsilon ^4 \, , & \nn
f_{17} &= \mathcal{T}_{17} \epsilon ^3 \, , &
f_{18} &= \mathcal{T}_{18} \epsilon ^3 \, , &
f_{19} &= \mathcal{T}_{19} \epsilon ^3 \, , &
f_{20} &= \mathcal{T}_{20} \epsilon ^3 \, , & \nn
f_{21} &= \mathcal{T}_{21} \epsilon ^4 \, , &
f_{22} &= \mathcal{T}_{22} \epsilon ^4 \, , &
f_{23} &= \mathcal{T}_{23} \epsilon ^4 \, , &
f_{24} &= \mathcal{T}_{24} \epsilon ^4 \, , & \nn
f_{25} &= \mathcal{T}_{25} \epsilon ^5 \, , &
f_{26} &= \mathcal{T}_{26} \epsilon ^4 \, , &
f_{27} &= \mathcal{T}_{27} \epsilon ^5 \, , &
f_{28} &= \mathcal{T}_{28} \epsilon ^4 \, , & \nn
f_{29} &= \mathcal{T}_{29} \epsilon ^4 \, , &
f_{30} &= \mathcal{T}_{30} \epsilon ^4 \, , &
f_{31} &= \mathcal{T}_{31} \epsilon ^5 \, , &
f_{32} &= \mathcal{T}_{32} \epsilon ^4 \, , & \nn
f_{33} &= \mathcal{T}_{33} \beta_{-} \epsilon ^4 \, , &
f_{34} &= \mathcal{T}_{34} \epsilon ^4 \, , &
f_{35} &= \mathcal{T}_{35} \epsilon ^5 \, , &
f_{36} &= \mathcal{T}_{36} \epsilon ^5 \, , & \nn
f_{37} &= \mathcal{T}_{37} \epsilon ^5 \, , &
f_{38} &= \mathcal{T}_{38} \epsilon ^5 \, , &
f_{39} &= \mathcal{T}_{39} \epsilon ^4 \, , &
f_{40} &= \mathcal{T}_{40} \epsilon ^5 \, , & \nn
f_{41} &= \mathcal{T}_{41} \epsilon ^3 \beta_{+} \, , &
f_{42} &= \mathcal{T}_{42} \beta_{-} \epsilon ^4 \, , &
f_{43} &= \mathcal{T}_{43} \epsilon ^5 \, , &
f_{44} &= \mathcal{T}_{44} \epsilon ^3 \beta_{+} \, , & \nn
f_{45} &= \mathcal{T}_{45} \epsilon ^5 \, , &
f_{46} &= \mathcal{T}_{46} \epsilon ^5 \, , &
f_{47} &= \mathcal{T}_{47} \epsilon ^6 \, , &
f_{48} &= \mathcal{T}_{48} \epsilon ^5 \, , & \nn
f_{49} &= \mathcal{T}_{49} \epsilon ^6 \, , &
f_{50} &= \mathcal{T}_{50} \epsilon ^6 \, , &
f_{51} &= \mathcal{T}_{51} \epsilon ^5 \, , &
f_{52} &= \mathcal{T}_{52} \beta_{-} \epsilon ^5 \, , & \nn
f_{53} &= \mathcal{T}_{53} \epsilon ^5 \, , &
f_{54} &= \mathcal{T}_{54} \epsilon ^5 \, , &
f_{55} &= \mathcal{T}_{55} \epsilon ^6 \, , &
f_{56} &= \mathcal{T}_{56} \epsilon ^5 \, , & \nn
f_{57} &= \mathcal{T}_{57} \epsilon ^5 \, , &
f_{58} &= \mathcal{T}_{58} \epsilon ^6 \, , &
f_{59} &= \mathcal{T}_{59} \epsilon ^5 \, , &
f_{60} &= \mathcal{T}_{60} \epsilon ^5 \, , & \nn
f_{61} &= \mathcal{T}_{61} \epsilon ^5 \, , &
f_{62} &= \mathcal{T}_{62} \epsilon ^5 \, , &
f_{63} &= \mathcal{T}_{63} \epsilon ^4 \, , &
f_{64} &= \mathcal{T}_{64} \epsilon ^4 \, , & \nn
f_{65} &= \mathcal{T}_{65} \epsilon ^3 \beta_{+} \, , &
f_{66} &= \mathcal{T}_{66} \epsilon ^4 \, , &
f_{67} &= \epsilon ^3 \left(\mathcal{T}_{67} \beta_{+}-2 \mathcal{T}_{59} \epsilon \right) \, , &
f_{68} &= \mathcal{T}_{68} \epsilon ^6 \, , & \nn
f_{69} &= \mathcal{T}_{69} \epsilon ^5 \, , &
f_{70} &= \mathcal{T}_{70} \beta_{-} \epsilon ^5 \, , &
f_{71} &= \mathcal{T}_{71} \epsilon ^5 \, , &
f_{72} &= \mathcal{T}_{72} \epsilon ^6 \, , & \nn
f_{73} &= \mathcal{T}_{73} \epsilon ^5 \, , &
f_{74} &= \mathcal{T}_{74} \epsilon ^5 \, , &
f_{75} &= \mathcal{T}_{75} \epsilon ^5 \, , &
f_{76} &= \mathcal{T}_{76} \beta_{-} \epsilon ^5 \, , & \nn
f_{77} &= \mathcal{T}_{77} \epsilon ^5 \, , &
f_{78} &= \mathcal{T}_{78} \epsilon ^6 \, , &
f_{79} &= \mathcal{T}_{79} \epsilon ^6 \, , &
f_{80} &= \mathcal{T}_{80} \epsilon ^6 \, , & \nn
f_{81} &= \mathcal{T}_{81} \epsilon ^6 \, , &
f_{82} &= \mathcal{T}_{82} \epsilon ^6 \, , &
f_{83} &= \mathcal{T}_{83} \epsilon ^6 \, , &
f_{84} &= \mathcal{T}_{84} \epsilon ^6 \, , &
\end{align}
where $f_{67}$ receives contributions from $\mathcal{T}_{59}$. Similar to the approach used for the previous integral family, the new basis $\mathbf{f}$ satisfies a differential equation that is linear in $\epsilon$.
By incorporating the leading singularities and performing an additional transformation obtained through the Magnus expansion, we successfully arrive at the canonical basis, as shown below.
\begin{align}
    F_1 &= f_1 s \, , &
F_2 &= f_2 m^2 \, , &
F_3 &= f_3 t \, , & \nn
F_4 &= f_4 m^2 t \, , &
F_5 &= f_5 t^2 \, , &
F_6 &= f_6 m^4 \, , & \nn
F_7 &= f_7 m^2 t \, , &
F_8 &= f_8 \left(m^2-t\right) \, , &
F_9 &= f_9 \left(m^2-t\right) \, , & \nn
F_{10} &= f_{10} \lambda \, , &
F_{11} &= f_{11} m^4 + 2f_{10} \left(s-2m^2\right)\, , &
F_{12} &= f_{12} \left(m^2-t\right) \, , & \nn
F_{13} &= f_{13} \left(m^2-t\right) \, , &
F_{14} &= f_{14} m^2 t-4 f_{13} m^2 \, , &
F_{15} &= f_{15} t \, , & \nn
F_{16} &= f_{16} \lambda \, , &
F_{17} &= f_{17} m^4 +2f_{16} \left(s-2m^2\right)\, , &
F_{18} &= f_{18} m^2 t^2 \, , & \nn
F_{19} &= f_{19} t^3 \, , &
F_{20} &= f_{20} m^4 t \, , &
F_{21} &= f_{21} t \left(t-m^2\right) \, , & \nn
F_{22} &= f_{22} m^2 \left(m^2-t\right) \, , &
F_{23} &= f_{23} t \left(t-m^2\right) \, , &
F_{24} &= f_{24} m^2 \left(m^2-t\right) \, , & \nn
F_{25} &= f_{25} \left(m^2-s-t\right) \, , &
F_{26} &= f_{26} s t \, , &
F_{27} &= f_{27} \left(m^2-s-t\right) \, , & \nn
F_{28} &= f_{28} s t \, , &
F_{29} &= f_{29} m^2 s \, , &
F_{30} &= f_{30} s t \, , & \nn
F_{31} &= f_{31} \left(m^2-s-t\right) \, , &
F_{32} &= f_{32} s t \, , &
F_{33} &= f_{33} \left(m^2-t\right) \, , & \nn
F_{34} &= f_{34} s t \, , &
F_{35} &= f_{35} \left(m^2-t\right) \, , &
F_{36} &= f_{36} \lambda \, , & \nn
F_{37} &= f_{37} \left(m^2-t\right) \, , &
F_{38} &= f_{38} \left(m^2-t\right) \, , &
F_{39} &= f_{39} m^2 \left(m^2-t\right) \, , & \nn
F_{40} &= f_{40} \left(m^2-t\right) \, , &
F_{41} &= f_{41} m^2 t-12 f_{40} m^2 \, , &
F_{42} &= f_{42} t \, , & \nn
F_{43} &= f_{43} \lambda \, , &
F_{44} &= 
f_{44} m^4
+3f_{43} \left(\frac{3s}{2}-4 m^2\right)
-\frac{f_{10} s}{2} \, , &
F_{45} &= f_{45} t \left(t-m^2\right) \, , & \nn
F_{46} &= f_{46} m^2 \left(m^2-t\right) \, , &
F_{47} &= f_{47} s \, , &
F_{48} &= f_{48} \left(m^2-t\right)^2 \, , & \nn
F_{49} &= f_{49} \left(m^2-t\right) \, , &
F_{50} &= f_{50} \left(m^2-s-t\right) \, , &
F_{51} &= f_{51} \left(s t-m^2 t+m^4\right) \, , & \nn
F_{52} &= f_{52} \left(m^2-t\right) \, , &
F_{53} &= f_{53} \left(s t-m^2 t+m^4\right) \, , &
F_{54} &= f_{54} \left(s t-m^2 t+m^4\right) \, , & \nn
F_{55} &= f_{55} \left(m^2-t\right) \, , &
F_{56} &= f_{56} t \left(t-m^2\right) \, , &
F_{57} &= f_{57} \left(s t-m^2 t+m^4\right) \, , & \nn
F_{58} &= f_{58} \lambda \, , &
F_{59} &= f_{59} m^2 \left(m^2-t\right) \, , &
F_{60} &= f_{60} \left(m^2-t\right)^2 \, , & \nn
F_{61} &= 
\rlap{$\displaystyle
f_{61} \left(s t-m^2 t+m^4\right) \, , \qquad \qquad
F_{62} = f_{62} m^2 \left(m^2-s-t\right)\, ,
$}
 & \nn
F_{63} &= 
\rlap{$\displaystyle
f_{63} m^2 t \left(t-m^2\right)
+f_{61} m^2 \left(m^2-t\right)
+2 f_{60} m^2 \left(m^2-t\right)
 \, , 
$}
& \nn
F_{64} &= f_{64} m^4 s \, , &
F_{65} &= f_{65} m^4 t-12 f_{59} m^4 \, , &
F_{66} &= f_{66} m^4 \left(m^2-t\right) \, , & \nn
F_{67} &= 
\rlap{$\displaystyle
\frac{f_{67} \lambda  m^4}{s}
-\frac{2 f_{66} \lambda  m^4 \left(m^2-t\right)}{s}
-f_{64} \lambda  m^4
+\frac{f_{63} \lambda  m^2 t \left(m^2-t\right)}{s}
-f_{61} \lambda  t $}
& \nn
& \quad \rlap{$\displaystyle
-\frac{2 f_{60} \lambda  m^2 \left(m^2-t\right)}{s}
+\frac{2 f_{59} \lambda  m^4}{s}
-\frac{2 f_{44} \lambda  m^4}{s}
-\frac{9 f_{43} \left(s-4 m^2\right)^2}{2 \lambda }
-\frac{9 f_{39} \lambda  m^2 \left(m^2-t\right)}{s}$}
& \nn
& \quad \rlap{$\displaystyle
+\frac{24 f_{38} \lambda  \left(m^2-t\right)}{s}
+6 f_{29} \lambda  m^2
+f_{28} \lambda  t
-\frac{12 f_{27} \lambda  \left(-m^2+s+t\right)}{s}
-\frac{6 f_{24} \lambda  m^2 \left(m^2-t\right)}{s}$}
& \nn
& \quad \rlap{$\displaystyle
+\frac{f_{10} \lambda  \left(6 m^2-s\right)}{s}
-\frac{8 f_{15} \lambda  t}{s}
-\frac{3 f_{11} \lambda  m^4}{2 s}
+\frac{20 f_8 \lambda  \left(m^2-t\right)}{s}
+\frac{3 f_7 \lambda  m^2 t}{s}
+\frac{2 f_2 \lambda  m^2}{s}
+f_1 \lambda 
\, , $}
& \nn
F_{68} &= 
\rlap{$\displaystyle
f_{68} \left(s t+t^2-2 m^2 t+m^4\right) \, ,  \qquad 
F_{69} = f_{69} s t^2 \, ,  \qquad 
F_{70} = f_{70} t \left(t-m^2\right) 
\, ,$}
 & \nn
F_{71} &= f_{71} s t^2 \, , &
F_{72} &= f_{72} t \left(m^2-s-t\right) \, , &
F_{73} &= f_{73} s t^2 \, , & \nn
F_{74} &= f_{74} m^2 s t \, , &
F_{75} &= f_{75}  m^2 t \left(t-m^2\right) \, , &
F_{76} &= f_{76} \left(m^2-t\right)^2 \, , & \nn
F_{77} &= f_{77} s t^2 \, , &
F_{78} &= f_{78} \lambda  t \, , &
F_{79} &= f_{79} t \left(s t-m^2 t+m^4\right) \, , & \nn
F_{80} &= 
\rlap{$\displaystyle
f_{80} t \left(t-m^2\right) 
+f_{79} m^2 t \left(m^2-t\right)
+f_{53} m^2 \left(m^2-t\right)
+\frac{1}{2} f_{51} m^2 \left(t-m^2\right)
\, , $}
& \nn
F_{81} &= 
\rlap{$\displaystyle
f_{81} t \left(t-m^2\right) 
+f_{57} m^2 \left(t-m^2\right)
+\frac{1}{2} f_{54} m^2 \left(m^2-t\right)
+\frac{1}{2} f_{51} m^2 \left(m^2-t\right)
\, ,$}
 & \nn
F_{82} &= f_{82} s t^3 \, , &
F_{83} &= f_{83} t^2 \left(t-m^2\right)+f_{79} m^2 t \left(m^2-t\right) \, , & \nn
F_{84} &= 
\rlap{$\displaystyle
f_{84} t^2 \left(t-m^2\right)+f_{69} m^2 t \left(m^2-t\right)+f_{77} m^2 t \left(t-m^2\right) 
\, . $}
&
\label{eq: ut basis 2}
\end{align}
The basis $\bF$ in family $\mathrm{LB}$ also satisfies a differential equation in the $d\text{log}$ form, similar to Eq. (\ref{eq: dlog}). However, the letters differ slightly from those in Eq. (\ref{eq: letters1}); specifically, $\omega_8$ is absent, and a new symbol letter appears in its place,
\begin{align}
\label{eq: letters2}
\omega _1 &= x\, ,&
\omega _4 &= 1+x\, ,&
\omega _7 &= 1+x y\, ,& \nn
\omega _2 &= y\, ,&
\omega _5 &= 1-y\, ,&
\omega _8 &= 1+ x+x y +x^2\, ,& \\
\omega _3 &= 1-x\, ,&
\omega _6 &= x+y\, ,&
\omega _9 &= x+y(1+x+x^2)\, .& \nonumber
\end{align}
They also match the ones occurring in the two-loop calculations~\cite{Gehrmann:2013cxs, Gehrmann:2014bfa}.

\section{Boundary conditions and solutions}
\label{sec: soulution}

In this section, we outline the process for solving the canonical differential equations and determining the boundary conditions that finalize the solutions. As the procedures are similar for both families, we provide a general overview below. The canonical basis is expanded as a Taylor series,
\begin{equation}
    \bF(x,y) = \sum_{n=0}^{\infty} \bF^{(n)}(x,y) \; \epsilon^n.
\end{equation}
Thanks to the factorization of $\epsilon$ in the differential equation, one can immediately obtain
\begin{equation}
    {\rm d} \bF^{(n)} =
    \begin{cases}
        {\rm d} \mathbb{A} \bF^{(n-1)} & n>0 \\
        \qquad 0 & n=0
    \end{cases}.
\end{equation}
This implies that the leading-order term, $\bF^{(0)}(x, y)$, is simply a constant vector. For $n > 0$, the equation can be solved recursively. Given that all the symbol letters are linear in $y$, it is a reasonable choice to first integrate along the $y$ axis, followed by the $x$ integration.
As a result, the solution at $\epsilon^n$ reads
\begin{equation}
    \bF^{(n)}(x,y) = \bF^{(n)}_y(x,y) + \bF^{(n)}_x(x) + {\bf c}^{(n)},
    \label{eq: Fnxy}
\end{equation}
where $\bf c$ is a constant vector to be fixed by boundary conditions. For the $y$ integration, we have
\begin{equation}
    \bF^{(n)}_y(x,y) = \int_0^y \mathrm{A}_y(x, b) \bF^{(n-1)}(x, b) {\rm d} b.
    \label{eq: Fny}
\end{equation}
It is straightforward to express it with MPL of weight $n$,
\begin{equation}
	G(w_n,...,w_1;z)=\int_0^z \frac{1}{t-w_n}\,G(w_{n-1},...,w_1;t)\,{\rm d} t,
\end{equation}
with 
\begin{equation}
	G(w_1;z)=\int_0^z \frac{1}{t-w_1} dt ~~ w_1\neq 0, ~~~~
	G(\underbrace{0,...,0}_{n \text{ times}};z)=\frac{\log^n(z)}{n!}, ~~~~
        G(;z) = 1.
\end{equation}
The indices of MPL appearing in $\bF^{(n)}_y(x,y)$ belong to 
\begin{equation}
\begin{split}
    \mathrm{LA} &: \left\{0,1,-x,-\frac{1}{x},-\frac{x^2+1}{x},-\frac{x}{x^2+x+1}\right\}; \\
    \mathrm{LB} &: \left\{0,1,-x,-\frac{1}{x},-\frac{x^2+x+1}{x},-\frac{x}{x^2+x+1}\right\}.
\end{split}
\end{equation}
To calculate $\bF^{(n)}_x(x)$, let's derive the differential equation it satisfies,
\begin{equation}
    \partial_x \bF^{(n)}_x(x) = -\partial_x \bF^{(n)}_y(x,y) + \mathrm{A}_x(x,y) \bF^{(n-1)}(x,y).
\end{equation}
Using Eq. (\ref{eq: Fny}), we have
\begin{equation}
\begin{split}
    \partial_x \bF^{(n)}_y(x,y) 
    &= \int_0^y {\rm d} b \left[\partial_x \mathrm{A}_y(x,b) \bF^{(n-1)}(x,b) + \mathrm{A}_y(x,b) \partial_x \bF^{(n-1)}(x,b)\right] \\
    &= \int_0^y {\rm d} b \left[\partial_b \mathrm{A}_x(x,b) \bF^{(n-1)}(x,b) + \mathrm{A}_x(x,b) \mathrm{A}_y(x,b) \bF^{(n-2)}(x,b)\right] \\
    &= \int_0^y {\rm d} b \left[\partial_b \mathrm{A}_x(x,b) \bF^{(n-1)}(x,b) + \mathrm{A}_x(x,b) \partial_b \bF^{(n-1)}(x,b)\right] \\
    &= \int_0^y {\rm d} b \; \partial_b\left[ \mathrm{A}_x(x,b) \bF^{(n-1)}(x,b)\right],
\end{split}
\end{equation}
where we used the integrability condition
\begin{equation}
    \partial_x \mathrm{A}_y = \partial_y \mathrm{A}_x, \qquad [\mathrm{A}_x, \mathrm{A}_y] = 0.
\end{equation}
Then we simply have
\begin{equation}
    \partial_x \bF^{(n)}_x(x) = \mathrm{A}_x(x,0) \bF^{(n-1)}(x,0),
\end{equation}
where $\bF^{(n-1)}(x,0)$ is understood as setting all the MPLs with $y$ being the argument to zero~\footnote{Namely, the MPLs in $\bF^{(n-1)}(x,0)$ are taken as their \textit{regularized} versions~\cite{Goncharov:2001iea}.}.
As a result,
\begin{equation}
    \bF^{(n)}_x(x) = \int_0^x {\rm d} a \mathrm{A}_x(a,0) \left[\bF^{(n-1)}_x(a) + {\bf c}^{(n-1)}\right],
\end{equation}
which again evaluates to MPLs of wight $n$ with indices being the elements of the set
\begin{equation}
\begin{split}
    \mathrm{LA} &: \left\{-1, 0, 1, -i, i, e^{-2i\pi/3}, e^{2i\pi/3}\right\}; \\
    \mathrm{LB} &: \left\{-1, 0, 1, e^{-2i\pi/3}, e^{2i\pi/3}\right\}. \\
\end{split}
\end{equation}
We solve the equation up to $\epsilon^6$, involving MPLs through weight six. It is worth noting that at two loops~\cite{Gehrmann:2013cxs}, the indices $-\frac{1+x^2}{x}$ ($\pm i$) of MPLs with argument $y$ ($x$) did not appear. We observe that these indices appear only at weight six and specifically for $F_{78}$ in family $\mathrm{LA}$.

Finally, to fix the undetermined integration constants, we use the regularity of all the canonical MIs in the following limits,
\begin{equation}
\label{eq: regularity}
    t \rightarrow m^2, \quad
    s \rightarrow -\frac{(t-m^2)^2}{t}, \quad
    s \rightarrow 2m^2 - t, \quad
    s \rightarrow m^2 - t.
\end{equation}
These constraints yield 81 (78) independent linear relationships among the 94 (84) boundary constants for families $\mathrm{LA}$ ($\mathrm{LB}$) at each order in $\ep$. The remaining 13 (6) constants must be determined through alternative independent methods in $\mathrm{LA}$ ($\mathrm{LB}$). Notably, these remaining constants are sufficiently simple to be calculated via direct integration with the assistance of the package {\sc HyperInt}~\cite{Panzer:2014caa}. For family $\mathrm{LA}$, the 13 input integrals are given as follows,
\begin{align}
\label{eq: input MI}
F_1 &= v^{-3\ep} \; \sigma_1
\,, &
F_2 &=  \; \sigma_1
\,, &
F_4 &= v^{-2\ep} \; \sigma_2
\,, &
F_5 &= v^{-3\ep} \; \sigma_2
\,, \nn
F_6 &= v^{-\ep} \; \sigma_2
\,, &
F_{16} &= v^{-3\ep} \; \sigma_3
\,, &
F_{17} &= v^{-3\ep} \; \sigma_4
\,, &
F_{18} &= v^{-3\ep} \; \sigma_5
\,, \nn
F_{19} &= -v^{-2\ep}\,, &
F_{20} &= -v^{-3\ep}\,, &
F_{22} &= v^{-2\ep} \; \sigma_6
\,, &
F_{23} &= v^{-3\ep} \; \sigma_6
\,, \nn
F_{37} &= v^{-3\ep} \; \sigma_7
\,, 
\end{align}
and as for family $\mathrm{LB}$, the needed integrals read
\begin{align}
F_1 &= v^{-3\ep} \; \sigma_1
\,, &
F_2 &=  \; \sigma_1
\,, &
F_4 &= y^{-\ep} \; \sigma_2
\,, \nn
F_{15} &= y^{-3\ep} \; \sigma_3
\,, &
F_{18} &= -y^{-2\ep}\,, &
F_{42} &= y^{-3\ep} \; \sigma_7
\,. 
\end{align}
In above equations, we used
\begin{align}
    \sigma_1 &= -1+22 \zeta _3 \epsilon ^3+33 \zeta _4 \epsilon ^4+234 \zeta _5 \epsilon ^5+\left(530 \zeta _6-242 \zeta _3^2\right) \epsilon ^6+O\left(\epsilon ^7\right), \nn
    \sigma_2 &= -1+6 \zeta _3 \epsilon ^3+9 \zeta _4 \epsilon ^4+42 \zeta _5 \epsilon ^5+\left(90 \zeta _6-18 \zeta _3^2\right) \epsilon ^6+O\left(\epsilon ^7\right), \nn
    \sigma_3 &= -\frac{1}{3}+\frac{16 \zeta _3 \epsilon ^3}{3}+8 \zeta _4 \epsilon ^4+64 \zeta _5 \epsilon ^5+\left(\frac{440 \zeta _6}{3}-\frac{128 \zeta _3^2}{3}\right) \epsilon ^6+O\left(\epsilon ^7\right), \nn
    \sigma_4 &= \frac{1}{6}+\frac{\zeta _2 \epsilon ^2}{2}-\frac{8 \zeta _3 \epsilon ^3}{3}-\frac{7 \zeta _4 \epsilon ^4}{8}-\frac{4}{3} \left(6 \zeta _2 \zeta _3+24 \zeta _5\right) \epsilon ^5+\left(\frac{64 \zeta _3^2}{3}-\frac{7775 \zeta _6}{96}\right) \epsilon ^6+O\left(\epsilon ^7\right), \nn
    \sigma_5 &= \frac{1}{9}+\frac{2 \zeta _2 \epsilon ^2}{3}-\frac{16 \zeta _3 \epsilon ^3}{9}+4 \zeta _4 \epsilon ^4-\frac{16}{9} \left(6 \zeta _2 \zeta _3+12 \zeta _5\right) \epsilon ^5+\left(\frac{128 \zeta _3^2}{9}-\frac{335 \zeta _6}{9}\right) \epsilon ^6+O\left(\epsilon ^7\right), \nn
    \sigma_6 &= \frac{1}{4}+\frac{\zeta _2 \epsilon ^2}{2}-\zeta _3 \epsilon ^3+\left(-2 \zeta _2 \zeta _3-9 \zeta _5\right) \epsilon ^5+\left(2 \zeta _3^2-\frac{91 \zeta _6}{4}\right) \epsilon ^6+O\left(\epsilon ^7\right), \nn
    \sigma_7 &= -6 \zeta _3 \epsilon ^3-9 \zeta _4 \epsilon ^4-102 \zeta _5 \epsilon ^5+\left(78 \zeta _3^2-240 \zeta _6\right) \epsilon ^6+O\left(\epsilon ^7\right).
\end{align}
Those integration constants at each order are linear combinations of certain numbers summarized in the following Table \ref{tab: boundary constants Euclidean}.
\begin{table}[htbp]
    \centering
    \begin{tabular}{c|cccccc}
      weight    &  1 & 2 & 3 & 4 & 5 & 6\\
      number(s)  & 0 & $\zeta_2$ & $\zeta_3$ & $\zeta_4$ & $\zeta_2\zeta_3, \zeta_5$ & $\zeta_3^2, \zeta_6$
    \end{tabular}
    \caption{The appearing numbers in boundary constants of different weights in the Euclidean region. At weight one, ${\bf c}^{(1)}$ is a zero vector.}
    \label{tab: boundary constants Euclidean}
\end{table}
The complete solutions for all the canonical MIs in the two integral families, up to weight six, encompass a total of 4151 MPLs. These are distributed across each order as illustrated in Table \ref{tab: mpl distribution}. Notably, nearly half of these MPLs are of weight six. 
\begin{table}[htbp]
    \centering
    \begin{tabular}{c|cccccc}
      weight    &  1 & 2 & 3 & 4 & 5 & 6\\
     \# of MPLs  & 14 & 62 & 305 & 975 & 836 & 1959
    \end{tabular}
    \caption{The number of MPLs of different weights.}
    \label{tab: mpl distribution}
\end{table}

To validate our results, we perform a comparison against the numerical calculations provided by {\sc pySecDec}~\cite{Borowka:2017idc}. We choose a point in the non-physical region at $(s, t, m^2) = (-5 \text{ GeV}^2, -0.5 \text{ GeV}^2, -1 \text{ GeV}^2)$, which yields
\begin{equation}
    x = \frac{3 - \sqrt{5}}{2}, \qquad y = \frac{1}{2}.
\end{equation}
Using {\sc GiNaC}~\cite{Bauer:2000cp, Weinzierl:2002hv, Vollinga:2004sn, Vollinga:2005pk}, we evaluate the MPLs to obtain the analytic results for all the canonical MIs. We find a strong agreement for the relatively simpler MIs in both families. However, producing high-precision numbers for the more complex MIs with more than eight propagators remains a significant challenge for {\sc pySecDec}, even in the non-physical region. To illustrate this, we select two examples from the top sector of each family and present the comparison between the analytic evaluations and numerical integrations in Table \ref{tab: compare}.
\begin{table}[htbp]
\centering
\renewcommand{\arraystretch}{1.2}
\begin{tabular}{|c|cc|cc|}
\hline
\multirow{2}{*}{$\epsilon^n$} & \multicolumn{2}{c|}{$F_{90}$ of $\mathrm{LA}$} & \multicolumn{2}{c|}{$F_{82}$ of $\mathrm{LB}$} \\
\cline{2-5}
                  &  Analytic &  Numeric & Analytic  &  Numeric \\ \cline{1-5}
     $0$ & $1/36$           &  \makecell{ $0.0277777777$ \\ $\qquad 7777771(1\! \pm \!6)$ }  & $1/36$           & \makecell{$0.0277777777$ \\ $\qquad 7777772(7\! \pm \! 4)$} \\
     $1$ & $-0.01859529594$ & $-0.018595(33\! \pm \! 31)$         & $-0.01859529594$ & $-0.0185952(86\! \pm \! 13)$ \\
     $2$ & $1.022771339$    & $1.02277(4\! \pm \! 9)$             & $0.2003043060$   & $0.200304(5\! \pm \! 7)$ \\
     $3$ & $3.090338916$    & $3.090(30\! \pm \!17)$              & $-0.5393446288$  & $-0.53934(9\! \pm \! 9)$  \\
     $4$ & $23.38255043$    & $23.38(57\! \pm \!20)$              & $0.5887425058$   & $0.588(63\! \pm \! 10)$ \\
     $5$ & $128.2789603$    & $128.2(58\! \pm \!24)$              & $15.04327881$    & $15.04(36 \! \pm \! 10)$ \\
     $6$ & $646.9675830$    & $646.(91\! \pm \!25)$               & $41.31582962$    & $41.3(27 \! \pm \! 14)$ \\        
\hline
\end{tabular}
\caption{Comparison of two MIs in the top sector of family $\mathrm{LA}$ and $\mathrm{LB}$ at $(s, t, m^2)=(-5 \text{ GeV}^2, -0.5 \text{ GeV}^2, -1 \text{ GeV}^2)$.}
\label{tab: compare}
\end{table}

\section{Continuation to physical region}
\label{sec: continuation}

In the region defined by Eq. (\ref{eq: nonphy region 2}), all the MIs $F_i$ in both families are real-valued. The same holds true for the MPLs whose argument is $y$. Although individual MPLs with the argument $x$ may be complex, the imaginary parts cancel out when summing them to compute the MIs.

As stated at the end of Section \ref{sec: family}, to evaluate MPLs in the physical region, a positive infinitesimal imaginary part must be assigned to $y$, which is negative and belongs to the interval $(-1/x, -x)$. As a result, the MPLs with $y$ as an argument are no longer all real-valued; many of them will develop a non-vanishing imaginary part. 
In practice, one can assign a small imaginary part to $y$ in \textsc{GiNaC}. While this is a convenient choice, it is not ideal, as it may lead to poor performance in the numerical evaluation of MPLs and a loss of precision, which is closely tied to the accuracy of the small imaginary part. A better approach, which offers faster and more stable evaluations that are essential for applications to scattering amplitudes, would be to explicitly compute the real and imaginary parts of each individual MI. In this case, all the MPLs are real-valued.

To achieve this, we transform the solution into a different representation by applying the change of variables,
\begin{equation}
    y \rightarrow -z-x.
\end{equation}
This leads to the constraint
\begin{equation}
    0 < z < \frac{1}{x} - x,
\end{equation}
where $z$ carries a negative infinitesimal imaginary part in the physical region.
With this transformation, the letters of the two families become
\begin{align}
   \mathrm{LA} : \{\omega_i\} &= \Omega \cup \left\{1-x z\right\}; \\
   \mathrm{LB} : \{\omega_i\} &= \Omega \cup \left\{\left(1+x+x^2\right) z+(1+x) x^2\right\},
\end{align}
where
\begin{equation}
    \Omega = \{ 1-x,x,1+x,z,x+z,1+x+z,1+x-x z,1-x^2-x z \}.
\end{equation}

As a result, the indices of the MPLs whose argument is $z$ are drawn from
\begin{align}
    \mathrm{LA} :& \{0, -x, -1-x, \frac{1}{x}, \frac{1+x}{x}, \frac{1-x^2}{x}\}; \\
    \mathrm{LB} :& \{0, -x, -1-x, \frac{1+x}{x}, \frac{1-x^2}{x}, -\frac{x^2(1+x)}{1+x+x^2}\}.
\end{align}
It follows straightforwardly that MPLs with $z$ being the argument are real-valued. 
Another noteworthy outcome of this change of variables is that the indices of the MPLs with argument $x$ in both families do not contain any roots of unity; they are drawn from the set $ \{-1, 0, 1\} $. This indicates that the MPLs with $x$ as the argument are harmonic polylogarithms. Consequently, it becomes evident that every individual MPL is real-valued, as anticipated. 

There are two methods to re-express the MIs in terms of MPLs using the new variables. The first is to replace $y$ with $ -z - x $ and transform the MPLs to a fibration basis with respect to $z$ and $x$. This can be accomplished using the \texttt{ToFibrationBasis} function in the \textsc{PolyLogTools} package~\cite{Duhr:2019tlz}. However, this approach is not applicable if non-linear letters are encountered, as it does. Therefore, we found it more convenient to re-solve the differential equations using $x$ and $z$.
The integration constants are fixed directly in the physical region by using the same regularity conditions in Eq.~(\ref{eq: regularity}) and input integrals in Eq.~(\ref{eq: input MI}). In each order, they are linear combinations of certain numbers summarized in Table \ref{tab: boundary constants physical}.
\begin{table}[htbp]
    \centering
    \begin{tabular}{c|cccccc}
      weight    &  1 & 2 & 3 & 4 & 5 & 6\\
      number(s)  & $i\pi$ & $\zeta_2$ & $i\pi^3, \zeta_3$ & $i\pi\zeta_3, \zeta_4$ & $i\pi^5, \zeta_2\zeta_3, \zeta_5$ & $i\pi\zeta_5, i\pi^3\zeta_3, \zeta_3^2, \zeta_6$
    \end{tabular}
    \caption{The appearing numbers in boundary constants of different weights in the physical region.}
    \label{tab: boundary constants physical}
\end{table}
We observe that the total number of MPLs is slightly reduced (partially) due to the absence of spurious indices.

To perform a numerical check in the physical region, we adopt a different strategy, as it is quite challenging for \textsc{pySecDec} to produce accurate results at physical points. Instead, we use the solutions obtained in the previous section to calculate the MIs at a non-physical point and then numerically integrate the differential equation to a physical point using \textsc{DiffExp}~\cite{Hidding:2020ytt}. 
Then the resulting numbers are compared against the ones obtained by evaluating the analytic expressions. We find perfect agreement for several test points.

We provide the expressions for the canonical MIs in both families for the physical region in a \textsc{Mathematica}-readable format in~\cite{mygit}. Additionally, we include the results for two example points—one in the non-physical region and one in the physical region.
Furthermore, we also present an implementation of all the canonical MIs in the physical region using \textsc{GiNaC}. For more details, see Appendix~\ref{sec: app1} and~\ref{sec: app2}.

\section{Conclusion and outlook}
\label{sec: conclusion}

In this paper, we presented an analytic calculation of the three-loop four-point Feynman integrals with two off-shell legs that have identical masses. They will contribute to the triple-virtual corrections to diboson production at N3LO at the LHC. 
 We provided the analytic expressions for a total of 170 master integrals in terms of multiple polylogarithms, up to weight six, in both the Euclidean and physical regions. The results were numerically validated against \textsc{pySecDec} and \textsc{DiffExp} in both regions. Additionally, we implemented these master integrals into a \texttt{C++} code using \textsc{GiNaC} for the evaluation of multiple polylogarithms. This implementation is extendable, allowing for the computation of amplitudes when available, and can serve as a library for phenomenological studies.

Several directions for future investigation are worth highlighting. A key step forward is to compute the scattering amplitudes for diboson production at N3LO. The leading-color contributions could be a good start. These calculations will involve more complex diagrams, such as the so-called tennis-court diagrams, which pose additional challenges. Tackling these diagrams will be a reasonable next step. We anticipate that IBP reduction will be particularly challenging, and constructing canonical integrals will be a demanding task.

Another valuable direction involves simplifying the expressions of the master integrals. By simplification, we mean re-expressing the integrals in terms of alternative multiple polylogarithms, reducing their complexity, and improving the efficiency and stability of numerical evaluations. One approach to achieve this is through the use of the \textit{symbol technique}~\cite{goncharov2004galoissymmetriesfundamentalgroupoids, Goncharov:2010jf, Duhr:2011zq, Duhr:2012fh}, a powerful tool for uncovering non-trivial relations among multiple polylogarithms. An algorithm to derive functional equations among multiple polylogarithms was proposed in Ref.~\cite{Duhr:2011zq}. Promising results have shown up to weight four, although the complexity increases significantly at higher weights. We leave further exploration of these techniques to future work.

\acknowledgments
Ming-Ming Long would like to thank Yang Zhang for useful discussions and technical support at the very beginning of the project. 
This research was supported by the Deutsche Forschungsgemeinschaft (DFG, German Research Foundation) under grant 396021762 - TRR 257.

\appendix
\section{Implementation}
\label{sec: app1}

We have uploaded a tarball, \texttt{ladder.tar.gz}, along with our submission to \textsc{arXiv}~\footnote{A Git repository has also been created to maintain the implementation~\cite{mygit}.}. This tarball contains the implementation of the expressions for all the canonical MIs in the physical region. The code structure closely follows that presented in Ref.~\cite{Kreer:2021sdt}.
We utilized \textsc{FORM}~\cite{Ruijl:2017dtg} to generate highly optimized outputs, based on the algorithms in Refs.~\cite{Kuipers:2012tgm, Kuipers:2013pba, Ruijl:2013epa, Ruijl:2014spa}, for those complicated expressions. This optimization brought the compilation and evaluation processes under effective control.

To build the program, extract the tarball and enter the root directory. Before compiling, ensure that the \textsc{CMake} is supported and that the \textsc{GiNaC} library is accessible. The following commands will create an executable file named \texttt{ladder} in the \texttt{./bin} directory.
\begin{lstlisting}[language=bash]
  $ mkdir build
  $ cd build 
  $ cmake ..
  $ make -j$(nproc)
\end{lstlisting}
Once compiled, the user can run the program by executing
\begin{lstlisting}[language=bash]
  $ ./bin/ladder
\end{lstlisting}
in the root directory.
This will output the total of 170 canonical MIs at a specific physical point $(s=9 \text{ GeV}^2, t=-2 \text{ GeV}^2, m^2=1 \text{ GeV}^2)$ to the file \texttt{numeric\_master.m}. To evaluate the MIs at different points, modify the values of \texttt{s}, \texttt{t}, and \texttt{ms} in the script \texttt{ladder.cpp} accordingly. The variable \texttt{Digits}, a positive integer, controls the desired precision of the results.
Readers should note, however, that the evaluation of MPLs at weight six can be particularly time-consuming.

\section{Ancillary files}
\label{sec: app2}

We provide a detailed description of the ancillary materials available in the same repository as our implementation~\cite{mygit}. The ancillary files for the two integral families are organized into separate sub-directories: \texttt{./LA} and \texttt{./LB}. Each sub-directory contains the following files.
\begin{itemize}
    \item \texttt{UT.m} contains the definition of canonical MIs (cf. Eqs. (\ref{eq: ut basis 1}, \ref{eq: ut basis 2})). Each integral $\mathcal{T}_i$ is denoted as
    \begin{equation}
        \mathrm{LA/LB}[a_1,...,a_{15}].
    \end{equation}
    \item \texttt{dLogForm.m} contains the matrix $\mathbb{A}$ in the $d\log$ form (cf. Eq. (\ref{eq: dlog})).
    \item \texttt{anaMIG\_Euc.m, anaMIG\_Phy.m} contain the expressions of canonical MIs up to weight six in the Euclidean and physical regions, respectively.
    \item \texttt{numMIs\_Euc.m, numMIs\_Phy.m} contain the numeric results of canonical MIs at $(s, t, m^2)=(-5 \text{ GeV}^2, -0.5 \text{ GeV}^2, -1 \text{ GeV}^2)$ and $(s, t, m^2)=(9 \text{ GeV}^2, -2 \text{ GeV}^2, 1 \text{ GeV}^2)$, respectively.
    \item \texttt{kira\_config/integralfamily.yaml, kira\_config/kinematics.yaml} are the configuration files we used in \textsc{Kira}.
\end{itemize}

\bibliographystyle{JHEP}
\bibliography{references}

\end{document}